\documentclass
	[a4paper,singlecolumn,twoside,12pt,showkeys, prb]
	{revtex4-2}
\usepackage{amsmath,amssymb}	% AMS math fonts
\usepackage{bm}					% bold math
\usepackage{color}					% for color text and text box
\usepackage{graphicx}				% Include figure files
\usepackage{dcolumn}				% Align table columns on decimal point
\usepackage[colorlinks=true, citebordercolor={0 0 0}, linkcolor=blue, citecolor=blue, urlcolor=blue]{hyperref}
\usepackage{setspace}

\usepackage{siunitx}
\usepackage{nicefrac}

\usepackage{glossaries}
\usepackage{upgreek}

\newcommand{\figCapSkip}{\vspace{-4ex}}	% space between a figure and its caption
\makeatletter
\def\frontmatter@abstractwidth{0.9\textwidth}	% abstract width
\makeatother

\makeglossaries

\newglossaryentry{a_1}
{
    name=\ensuremath{a_1},
    description={Radius of the SRG rotor}
}

\newglossaryentry{rho}
{
    name=\ensuremath{\rho},
    description={Density of the SRG rotor}
}
\newglossaryentry{alpha}
{
    name=\ensuremath{\alpha},
    description={Thermal expansion coefficient of the SRG rotor}
}
\newglossaryentry{t}
{
    name=\ensuremath{t},
    description={Time}
}

\newglossaryentry{p}
{
    name=\ensuremath{p},
    description={Pressure}
}
\newglossaryentry{p_MC}
{
    name=\ensuremath{p_\mathrm{MC}},
    description={Monte Carlo truth value for the pressure}
}

\newglossaryentry{c_m}
{
    name=\ensuremath{c_m},
    description={Velocity slip coefficient}
}

\newglossaryentry{mu}
{
    name=\ensuremath{\mu},
    description={Viscosity}
}
\newglossaryentry{mu_MC}
{
    name=\ensuremath{\mu_\mathrm{MC}},
    description={Monte Carlo truth value for the viscosity}
}

\newglossaryentry{k_B}
{
    name=\ensuremath{k_B},
    description={Boltzmann constant}
}

\newglossaryentry{D}
{
    name=\ensuremath{D},
    description={Torque on the SRG rotor}
}

\newglossaryentry{I}
{
    name=\ensuremath{I},
    description={Moment of inertia of the SRG rotor}
}

\newglossaryentry{a_2}
{
    name=\ensuremath{a_2},
    description={Radius of the cylinder surrounding the SRG rotor}
}
\newglossaryentry{C_0}
{
    name=\ensuremath{C_0},
    description={Parameter describing the flow of a sphere rotating inside a cylinder with rotation axis perpendicular to the cylinder axis}
}
\newglossaryentry{T}
{
    name=\ensuremath{T},
    description={Temperature}
}

\newglossaryentry{Omega}
{
    name=\ensuremath{\Omega},
    description={The angular speed of the SRG rotor}
}
\newglossaryentry{omega}
{
    name=\ensuremath{\omega},
    description={The normalized angular deceleration of the SRG rotor}
}
\newglossaryentry{gamma}
{
    name=\ensuremath{\gamma},
    description={Coefficient of thermomolecular pressure difference}
}
\newglossaryentry{delta}
{
    name=\ensuremath{\delta},
    description={Rarefaction parameter}
}
\newglossaryentry{omega_MC}
{
    name=\ensuremath{\omega_\mathrm{MC}},
    description={The Monte Carlo value of the normalized angular deceleration of the SRG rotor}
}
\newglossaryentry{m}
{
    name=\ensuremath{m},
    description={The mass of the SRG rotor}
}
\newglossaryentry{M}
{
    name=\ensuremath{M},
    description={The molar mass of the investigated gas}
}

\begin{document}
%\setcounter{page}{\value{startPage}}

%%%%% DO NOT CHANGE THE ABOVE HERADERS %%%%%

%%%%% Several usefull commands %%%%%
%%%%% YOU CAN ADD YOUR OWN COMMANDS HERE %%%%%
\newcommand{\By}{$\times$}
\newcommand{\SqrtBy}[2]{$\sqrt{#1}$\kern0.2ex$\times$\kern-0.2ex$\sqrt{#2}$}
\newcommand{\Degree}{$^\circ$}
\newcommand{\DegreeC}{$^\circ$C}
\newcommand{\Ohmcm}{$\Omega\cdot$cm}

\title{%%%%% PUT TITLE OF THE PAPER HERE %%%%%
Viscosity measurements of gaseous H$_2$ between \SIrange{200}{300}{\kelvin} with a spinning rotor gauge
}

%%%%% PUT AUTHOR INFORMATIONS HERE %%%%%

\author{Johanna Wydra}
\email[Corresponding author: ]{johanna.wydra@kit.edu}
\affiliation{%
Institute for Astroparticle Physics - Tritium Laboratory Karlsruhe, Karlsruhe Institute for Technology (KIT), Hermann-von-Helmholtz-Platz 1, 76344 Eggenstein-Leopoldshafen, Germany}

\author{Alexander Marsteller}
%\altaffiliation{%
%Institute for Astroparticle Physics - Tritium Laboratory Karlsruhe, Karlsruhe Institute for Technology (KIT), Hermann-von-Helmholtz-Platz 1, 76344 Eggenstein-Leopoldshafen, Germany}
\affiliation{%
Department of Physics, University of Washington, Seattle, WA 98195, USA
}

\author{Robin Größle}
\affiliation{%
Institute for Astroparticle Physics - Tritium Laboratory Karlsruhe, Karlsruhe Institute for Technology (KIT), Hermann-von-Helmholtz-Platz 1, 76344 Eggenstein-Leopoldshafen, Germany}

\author{Michael Sturm}
\affiliation{%
Institute for Astroparticle Physics - Tritium Laboratory Karlsruhe, Karlsruhe Institute for Technology (KIT), Hermann-von-Helmholtz-Platz 1, 76344 Eggenstein-Leopoldshafen, Germany}

\author{Florian Priester}
\affiliation{%
Institute for Astroparticle Physics - Tritium Laboratory Karlsruhe, Karlsruhe Institute for Technology (KIT), Hermann-von-Helmholtz-Platz 1, 76344 Eggenstein-Leopoldshafen, Germany}

\author{Simon Gentner}
\affiliation{%
Institute for Astroparticle Physics - Tritium Laboratory Karlsruhe, Karlsruhe Institute for Technology (KIT), Hermann-von-Helmholtz-Platz 1, 76344 Eggenstein-Leopoldshafen, Germany}

\author{Linus Schlee}
\affiliation{%
Institute for Astroparticle Physics - Tritium Laboratory Karlsruhe, Karlsruhe Institute for Technology (KIT), Hermann-von-Helmholtz-Platz 1, 76344 Eggenstein-Leopoldshafen, Germany}

\begin{abstract}

\vspace*{1mm}

%%%%% PUT ABSTRACT HERE %%%%%
Experimental values for the viscosity of the radioactive hydrogen isotopologue tritium are still unknown in literature. 
Existing values from ab initio calculations disregard quantum mechanic effects and are therefore only good approximations for room temperature and above. 
To fill in these missing experimental values, a measurement setup has been designed, to measure the viscosity of gaseous hydrogen and its isotopologues (H$_2$, HD, HT, D$_2$, DT, T$_2$) at cryogenic temperatures.
In this paper, the first results with this Cryogenic Viscosity Measurement Apparatus (Cryo-ViMA) of the viscosity of gaseous hydrogen between \SIrange{200}{300}{\kelvin} are presented.

\end{abstract}

\keywords{%%%%% PUT KEYWORDS HERE %%%%%
% 3-to-5 keywords are required.
Viscosity measurements; hydrogen; Tritium Laboratory Karlsruhe; spinning rotor gauge; cryogenic setup
}

%%%%%%%%%% DO NOT CHANGE THE FOLLOWING %%%%%%%%%%
%\date[]{Received \receivedData; Accepted \acceptedData; Published \pulishedData} 
\maketitle
%\thispagestyle{ejssntFP}
%\pagestyle{ejssnt}
%%%%%%%%%% DO NOT CHANGE THE ABOVE %%%%%%%%%%
\newpage

%%%%%%%%%%%%%%%%%%%%%%
%% BODY OF THE PAPER
%%%%%%%%%%%%%%%%%%%%%%

%%%%%%%%%%%%%%%%%%%%%%
\section{Introduction} \label{sec:Intro}
%%%%%%%%%%%%%%%%%%%%%%
The viscosity of a gas is a fundamental material property important for gas dynamics calculations.
Measurements of gas viscosities have been done for a wide range of substances over a wide temperature range \cite{KESTIN19591033, Kestin1980, Rietveld1957, Itterbeek1938, Itterbeek1940, assael2018}.
Similarly, ab initio calculations have been performed for many simple systems such as the noble gases \cite{Hurly2000} or molecular hydrogen \cite{mehl2010, Schaefer2010, Song2016}. 
A simple molecule that has not been covered widely due to its radioactivity is the superheavy isotope of hydrogen, tritium.
While some ab initio calculations exist \cite{Song2016}, they have been carried out using a classical approach and neglect quantum effects, which makes them only suitable for high temperature of \SI{300}{\kelvin} and above.
To the best of the authors' knowledge, no experimental values for the viscosity of tritium exist in published literature.

However, the cryogenic region of viscosity covered by neither theoretical calculations nor experiment is of interest for several applications such as nuclear fusion or experimental neutrino physics.
The Karlsruhe Tritium Neutrino experiment (KATRIN) is one such experiment that aims to measure the electron antineutrino mass via spectroscopy of tritum $\upbeta$-decay electrons.
It has recently managed to set a new upper limit on the neutrino mass of \SI{0.8}{\electronvolt} (\SI{90}{\percent C.L.}) \cite{aker2022}.
In this experiment, tritium is circulated through a \SI{10}{\meter} long windowless gaseous tritium source (WGTS) at a temperature of \SI{80}{\kelvin}.
The density profile of tritium inside the WGTS is dependant on the viscosity of tritium, making its value of interest for the KATRIN experiment. 

In order to measure the viscosity of tritium down to temperatures relevant to the KATRIN experiment, we have built a viscosity measurement apparatus (ViMA) based on a spinning rotor gauge.
This apparatus is capable of measuring gas viscosities over a wide range of temperatures from \SIrange{77}{300}{\kelvin} in a manner that is compatible with the material requirements for later use with tritium.
In this paper, we report first measurements of the viscosity of hydrogen measured with this apparatus over a range of \SIrange{200}{300}{\kelvin}.

%%%%%%%%%%%%%%%%%%%%%%
\section{Measurement setup and procedure} \label{sec:SetupProcedure}
%%%%%%%%%%%%%%%%%%%%%%
\subsection{Measurement setup} \label{sec:SetupProcedure:subsec:Setup}
%%%%%%%%%%%%%%%%%%%%%%
\begin{figure}[tbh]
    \begin{center}
    \includegraphics[width=0.5\textwidth]{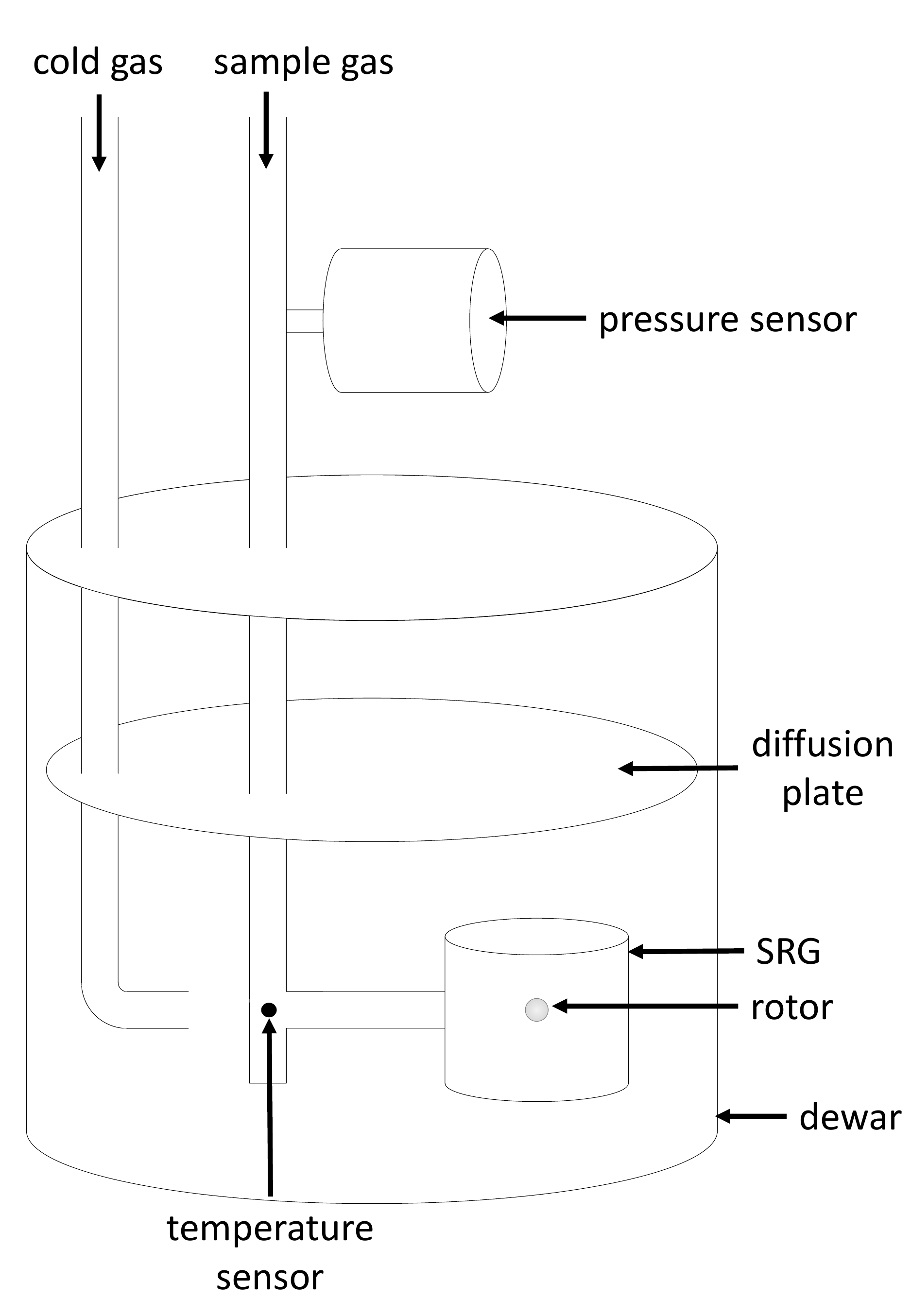}
    \end{center}\figCapSkip
    \caption{\label{fig:Prinzipbild}Simplifyed scetch of the Cryogenic Viscosity measurement apparatus (Cryo-ViMA).}
\end{figure}
The apparatus for measuring viscosities consists in principle of three main components: a spinning rotor gauge (SRG) (MKS SRG-3), a gas species independent membrane capacitance pressure sensor (MKS 626, \SI{20}{torr} full scale), and a Dewar cooled or heated by gaseous nitrogen.
The membrane pressure sensor at room temperature is connected to the same sample volume as the SRG, which is situated inside of the Dewar.
This sample volume can be evacuated using a turbo molecular pump and filled with different high purity gases (H2, D2, N2) and their mixtures.
All wetted surfaces of the setup are fully metal, ensuring compatibility for operation with tritium at a later point in time.
The Dewar cooling is provided by a temperature controllable gaseous nitrogen injection system which is fed from a liquid nitrogen Dewar (KGW-Isotherm TG-LKF H). 
This allows for operation of the experiment over a broad temperature range from \SI{45}{\celsius} down to cryogenic temperatures \SI{>77}{\kelvin}. 
A more detailed description of this setup can be found in \cite{Wydra2022unpublished}.
%%%%%%%%%%%%%%%%%%%%%%
\subsection{Measurement procedure} \label{sec:SetupProcedure:subsec:Procedure}
%%%%%%%%%%%%%%%%%%%%%%
A measurement with our setup consists of the simultaneous measurement of temperature of the SRG sample volume, the measured deceleration rate of the SRG, as well as of the pressure $\gls{p}$ measured by the gas species independent membrane capacitance sensor.
Measuring these values at a constant temperature $\gls{T}$ for varying values of $p$ allows for the calculation of the dynamic viscosity, which will be laid out in more detail below in \autoref{sec:DataAnalysis}.

From an experimental point of view, varying the pressure $p$ at constant temperature $T$, as it is done in \cite{Wydra2022}, is an untenable procedure for scanning a large range of temperatures.
Instead, we fill our sample volume with a certain pressure set point $p_S$, close off the volume, then cool down the system from ${\approx\!\SI{300}{\kelvin}}$ to ${\approx\!\SI{200}{\kelvin}}$ and let it warm up to ${\approx\!\SI{300}{\kelvin}}$ again. 
This procedure is repeated for several set points $p_S$.
Each such cycle takes around 1 day to complete, allowing for good temperature equilibration at any point in time.
The advantage of this procedure is, that the temperature inside the Dewar does not need to be stable at high level of \SI{0.1}{\kelvin\per\minute}. 
Since the temperature of the sample gas and of the nitrogen gas used for cooling is measured at different points inside the Dewar, it is only important, that the temperature inside the SRG does not change more than \SI{0.5}{\kelvin\per\minute}.
This can be easily achieved by the regulation of the cool-down-speed.
Furthermore, this approach allows for minimal need of operator intervention without a complex process control system as it would be necessary for automatic changing of the pressure.

%%%%%%%%%%%%%%%%%%%%%%
\section{Data analysis} \label{sec:DataAnalysis}
%%%%%%%%%%%%%%%%%%%%%%
The deceleration rate of the rotating sphere inside the SRG is dependant on the pressure and temperature of the sample gas. The viscosity of the gas is then determined by the following formula, derived in \cite{Tekasakul1996,Bentz1997,Loyalka1996}
\begin{equation}
    \frac{1}{\nicefrac{D}{\Omega}} = \frac{1}{8 \pi a_1^3 C_0 \mu} + \frac{1}{p} \cdot \sqrt{\frac{2 \gls{k_B} T}{m}} \left( \frac{\gls{c_m}}{8 \pi a_1^3 C_0} \left(\frac{3}{a_1} + \frac{1}{a_2}\right)\right)\,, \label{eq:srgvisc}
\end{equation}
where $\gls{D}$ is the torque on the sphere, $\gls{Omega}$ the angular speed of the sphere, $\gls{a_1}$ and $\gls{a_2}$ are the radii of the rotating sphere and the thimble surrounding it respectively.
The $\gls{C_0}$ is a calibration factor determined by a similar measurement with helium as a sample gas, where the viscosity $\gls{mu}$ is known to high precision from ab initio calculations\cite{Hurly2000}. 
The y-axis intercept of the linear fit in $\nicefrac{1}{p}$ of the data allows for the extraction of the viscosity $\mu$ given a known $C_0$ or vice versa.

The first results from such a setup at liquid nitrogen temperature can be found in \cite{Wydra2022}.
For the measurements with the cryogenic viscosity measurement apparatus (Cryo-ViMA), the procedure differs sightly from the previous measurements as described in \autoref{sec:SetupProcedure:subsec:Procedure}. 
Since the sample volume is closed off during a temperature cycle, the conserved quantity in a data-set from such a cycle is the number of particles.
Conversely, the pressure inside the SRG will decrease with decreasing temperature and vice versa. 
Since the pressure is logged away together with the temperature and the deceleration rate, the data analysis is not affected by this.
For each thermal cycle there exists one data-set, where all the data needed is logged every \SI{2}{\second}.
During the analysis, the data-sets are aligned according to temperature instead of number of particles. 
These data aligned in this manner are then used for the fit of \autoref{eq:srgvisc}.
This is done for all the temperatures found and the results can then be plotted as a temperature dependant viscosity, shown in \autoref{fig:Results} in \autoref{sec:Results}.

%%%%%%%%%%%%%%%%%%%%%%
\subsection{Temperature corrections} \label{sec:DataAnalysis:subsec:TCorrection}
%%%%%%%%%%%%%%%%%%%%%%
Previous measurements have shown, that the temperature has the highest impact on the uncertainty of the results. 
There are two major corrections done to the logged temperature values.
The first is a linear calibration correction to account for deviations of our PT100 sensor from the standard PT100 curve.
The second effect is an intrinsic effect of the SRG and is not trivial to correct for. 
The SRG is primarily a pressure gauge for ultra high vacuum (UHV). 
Under these conditions, the largest thermal change of the system is caused by the initial acceleration of the rotating sphere. 
There it is recommended to wait for at least \SI{30}{\minute}, to give the sphere the possibility to come into thermal equilibrium with the surroundings again \cite{MKSManual}. 
For the measurement of the viscosity in our setup this is not possible, since the deceleration rate is too high.
During our measurements, the sphere is re-accelerated every \SI{\approx 10}{\second} from \SIrange{420}{440}{\hertz}. 
With the deceleration rate and the thermal expansion of the sphere the heat of the sphere can be calculated \cite{jousten2018}.
With a basic finite elements method (FEM) simulation, done with Ansys\textsuperscript{\textregistered} Workbench 2020 R2, the temperature gradient of the gas surrounding the sphere can be calculated. 
Since this gradient is not only dependant on the sample gas but also on its density and the temperature, which both affect the deceleration rate, it has to be run for all possible combinations.
First simulations show a temperature gradient of up to \SI{3}{\kelvin} over a distance of \SI{3}{\milli\meter}, which is still under investigation.
Nevertheless our temperature sensor is not close enough to be able to measure this effect, so the measured temperature will likely be systematically lower than the true temperature around the sphere. 

%%%%%%%%%%%%%%%%%%%%%%
\subsection{Pressure corrections} \label{sec:DataAnalysis:subsec:pCorrection}
%%%%%%%%%%%%%%%%%%%%%%
In this setup the pressure is measured at room temperature while the gas temperature varies between \SIrange{200}{300}{\kelvin}, leading to a thermomolecular pressure difference.
The influence of the temperature gradient on the accuracy of the pressure measurement can be found in \cite{sharipov1996}.
In this setup the uncertainty from not including this effect is negligible.

%%%%%%%%%%%%%%%%%%%%%%
\section{Results and Discussion} \label{sec:Results}
%%%%%%%%%%%%%%%%%%%%%%
\begin{figure}[tbh]
    \begin{center}
    \includegraphics[width=\textwidth]{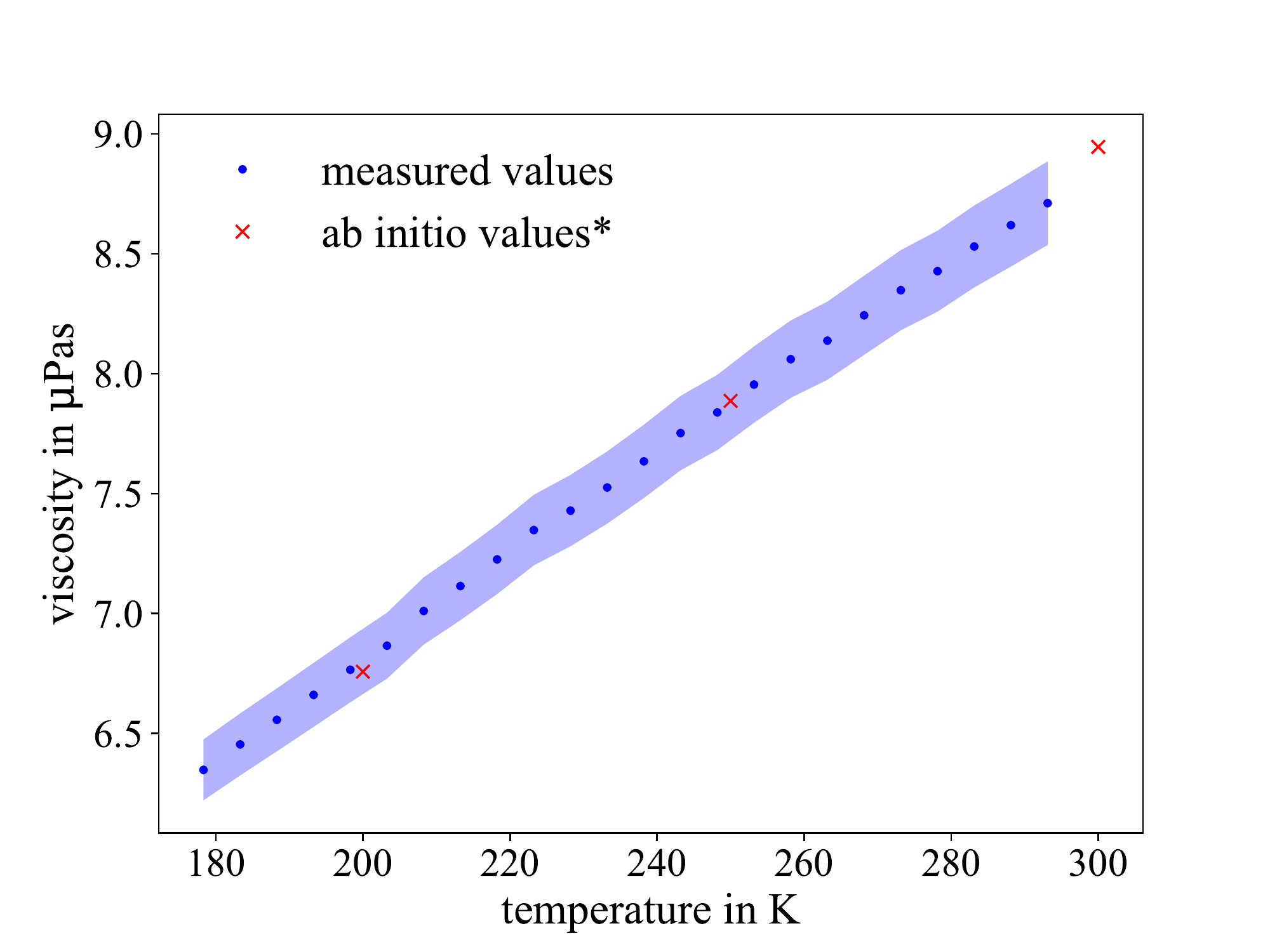}
    \end{center}\figCapSkip
    \caption{\label{fig:Results}Viscosity of hydrogen in dependence of the temperature.
    The blue dots are extracted data points, while the blue band shows the \SI{2}{\percent} uncertainty, obtained from previous measurements. *Literature values (red crosses) are taken from \cite{mehl2010}. }
\end{figure}
In \autoref{fig:Results} the results of the first measurements with Cryo-ViMA are shown. 
The measured values agree with the literature values within an uncertainty of about \SI{2}{\percent}.
The measured values show some slight tilt compared to the ab initio values by \cite{mehl2010}, which might be caused by the temperature gradient inside the SRG.
For further measurements it is planned to investigate the influence of the temperature gradient inside the SRG on the accuracy of the results.

%%%%%%%%%%%%%%%%%%%%%%
\section{Conclusion} \label{sec:Conclusion}
%%%%%%%%%%%%%%%%%%%%%%

We have presented first measurement results of our tritium compatible viscosity measurement apparatus on the viscosity of gaseous hydrogen in the temperature range between \SIrange{200}{300}{\kelvin}.
The results agree with ab initio calculated literature values \cite{mehl2010} within an uncertainty of \SI{2}{\percent}.
Known thermal effects which are suspected to be the main cause for the observed deviation are currently under investigation.
Based on the presented measurements, we are confident that after investigation and improvement of the observed uncertainties, Cryo-ViMA will  be able to measure the viscosity of tritium over a wide temperature range with an uncertainty of about \SI{1}{\percent}.

%%%%%%%%%%%%%%%%%%%%%%
\begin{acknowledgments}
We acknowledge the support of the Helmholtz Association (HGF), Germany, the German Ministry for Education and Research BMBF (05A20VK3) and the DFG graduate school KSETA, Germany (GSC 1085).
\end{acknowledgments}

\printglossary[title={List of Symbols}, nonumberlist]

%%%%%%%%%%%%%%%%%%%%%%
%% PUT LIST OF REFERENCES HERE
%%%%%%%%%%%%%%%%%%%%%%
%\vfill
% \bibliographystyle{prb}
\bibliography{bibliography}
%\begin{references}

%\end{references}

\end{document}